\documentclass[prd,preprint,superscriptaddress,amsmath,amssymb,nofootinbib]{revtex4}
\usepackage{graphicx}
\usepackage{dcolumn}
\usepackage{bm}
\usepackage{amssymb}
\usepackage{amsmath}
\usepackage{epsfig}    
\usepackage{color}
\usepackage{slashed}
\usepackage{hhline}

\def\be{\begin{equation}}
\def\ee{\end{equation}}
\newcommand{\bea}{\begin{eqnarray}}
\newcommand{\eea}{\end{eqnarray}}
\newcommand{\nn}{\nonumber}

\begin{document}

{\begin{flushright}{APCTP Pre2022 - 010}\end{flushright}}

\title{An alternative gauged $U(1)_R$ symmetric model\\
 in light of the CDF II $W$ boson mass anomaly}

\author{Keiko I. Nagao}
\email{nagao@ous.ac.jp}
\affiliation{Okayama University of Science, Faculty of Science,  Department of  Physics, Ridaicho 1-1, Okayama, 700-0005, Japan}

\author{Takaaki Nomura}
\email{nomura@scu.edu.cn}
\affiliation{College of Physics, Sichuan University, Chengdu 610065, China}

\author{Hiroshi Okada}
\email{hiroshi.okada@apctp.org}
\affiliation{Asia Pacific Center for Theoretical Physics (APCTP) - Headquarters San 31, Hyoja-dong,
Nam-gu, Pohang 790-784, Korea}
\affiliation{Department of Physics, Pohang University of Science and Technology, Pohang 37673, Republic of Korea}

\date{\today}

\begin{abstract}
 We consider an explanation of CDF II W bosom mass anomaly by $Z-Z'$ mixing with $U(1)_R$ gauge symmetry under which right-handed fermions are charged.
 It is found that $U(1)_R$ is preferred to be leptophobic to accommodate the anomaly while avoiding other experimental constraints.
 In such a case we require extra charged leptons to cancel quantum anomalies and the SM charged leptons get masses via interactions with the extra ones.
 These interactions also induce muon $g-2$ and lepton flavor violations.
 We discuss muon $g-2$, possible flavor constraints, neutrino mass generation via inverse seesaw mechanism, 
 and collider physics regarding $Z'$ production for parameter space explaining the W boson mass anomaly.
 
\end{abstract}
\maketitle

\section{Introduction}
Precision measurements of electroweak observable are good test of the  standard model(SM) and would provide a hint of beyond the SM.
CDFII collaboration recently reported updated result of the SM charged-gauge boson (W boson) mass~\cite{CDF:2022hxs}
\begin{align}
m_W = (80.433\pm 0.0064_{\rm stat}\pm 0.0069_{\rm syst})\ {\rm GeV},
\label{eq:Wmass}
\end{align}
which deviates from the SM prediction by 7$\sigma$, where the SM prediction indicates $m_W = (80.357\pm 0.006)\ {\rm GeV}$. 
The disagreement also appears to the previous global combination of data from LEP, CDF, D0, and ATLAS where they give the mass range of $m_W = (80.379\pm 0.012)\ {\rm GeV}$~\cite{ParticleDataGroup:2018ovx}.
This anomaly suggests new physics (NP) beyond the SM~\cite{Fan:2022dck, Lu:2022bgw, Athron:2022qpo, Yuan:2022cpw, Strumia:2022qkt, Yang:2022gvz, deBlas:2022hdk, diCortona:2022zjy, Du:2022pbp, Tang:2022pxh, Blennow:2022yfm, Cacciapaglia:2022xih, Sakurai:2022hwh, Fan:2022yly, Liu:2022jdq, Lee:2022nqz, Cheng:2022jyi, Song:2022xts, Bagnaschi:2022whn, Paul:2022dds, Bahl:2022xzi, Asadi:2022xiy, DiLuzio:2022xns, Athron:2022isz, Gu:2022htv, Heckman:2022the, Babu:2022pdn, Heo:2022dey, Du:2022brr, Cheung:2022zsb, Crivellin:2022fdf, Endo:2022kiw, Biekotter:2022abc, Balkin:2022glu, Krasnikov:2022xsi, Ahn:2022xeq, Han:2022juu, Zheng:2022irz, Kawamura:2022uft, Ghoshal:2022vzo, FileviezPerez:2022lxp, Nagao:2022oin, Kanemura:2022ahw, Mondal:2022xdy, Zhang:2022nnh, Borah:2022obi, Chowdhury:2022moc, Arcadi:2022dmt, Cirigliano:2022qdm, Carpenter:2022oyg, Popov:2022ldh, Ghorbani:2022vtv, Du:2022fqv, Bhaskar:2022vgk, Batra:2022org, Cao:2022mif, Zeng:2022lkk, Baek:2022agi, Borah:2022zim, Almeida:2022lcs, Cheng:2022aau, Heeck:2022fvl, Addazi:2022fbj, Lee:2022gyf, Cai:2022cti, Benbrik:2022dja, Yang:2022qgs, Batra:2022pej, Tan:2022bip, Abouabid:2022lpg, Chen:2022ocr, Zhou:2022cql, Gupta:2022lrt, Basiouris:2022wei, Wang:2022dte, Botella:2022rte, Barman:2022qix, Kim:2022hvh, Li:2022gwc, Isaacson:2022rts, Evans:2022dgq, Chowdhury:2022dps, Kim:2022zhj, Lazarides:2022spe, Senjanovic:2022zwy, Ghosh:2022zqs, Li:2022eby, Rodriguez:2022wix, Ramirez:2022zpk, Kawamura:2022fhm, Afonin:2022cbi, Allanach:2022bik, Xue:2022mde, Rizzo:2022jti, VanLoi:2022eir, YaserAyazi:2022tbn, Chakrabarty:2022voz, CentellesChulia:2022vpz,Kim:2022xuo}, and
can be interpreted as the deviation of  $\Delta T$ oblique parameter~\cite{Peskin:1991sw, Peskin:1990zt},
where the oblique parameters are zero in the SM.

One of the straightforward explanations of the anomaly can be realized by introducing extra $U(1)$ gauge symmetry 
where the SM Higgs field is charged under it.
Then $\Delta T$ is shifted by effect of mass mixing between the SM $Z$ and a new neutral gauge boson $Z'$.
In particular one of the minimal scenarios is the  explanation by $Z'$ with right-handed $U(1)_R$ symmetry~\cite{Jung:2009jz}.
The symmetry is originally  proposed by chiral  anomaly cancellations with three right-handed neutrinos, and it is well testable at the International Linear Collider (ILC) or the Large Hadron Collider(LHC) due to observing the difference of chirality~\cite{Nomura:2017tih, Nomura:2017abh, Nomura:2018mwr, Nomura:2017ezy, Chao:2017rwv,Seto:2020jal}.
However the W boson anomaly cannot be explained if charged-leptons have nonzero $U(1)_R$ charge because of the constraints from $W$ and $Y$ oblique parameters~\cite{Strumia:2022qkt}. Then, only the possibility to explain the W boson anomaly along this idea is that only the SM Higgs field and quarks have nonzero charge under $U(1)_R$ symmetry.
Note that another possibility is $U(1)_H$ case where only Higgs doublet is charged under it.
 In this case we need another Higgs doublet to induce the SM fermion masses and 
two Higgs doublets can also contribute to $\Delta T$ parameter at loop level.  
Remarkably in our setting $W$ boson anomaly is explained by purely $Z$--$Z'$ mixing effect. 

 In this letter, we extend the original $U(1)_R$ model to make the $Z'$ to be leptophobic in order to explain this anomaly avoiding other electroweak precision tests.
 Then extra $SU(2)$ singlet charged leptons are required to cancel quantum anomalies.
 As a result the masses of the SM charged leptons are obtained via interactions between the SM lepton and the extra charged leptons.
 Such interactions also induce lepton anomalous magnetic moments and lepton flavor violations (LFVs) at loop level~\cite{Lee:2021gnw,Guedes:2022cfy}.
 In addition, we include discussion of active Majorana neutrino mass matrix via inverse seesaw scenario~\cite{Nomura:2018mwr}.

 This letter is organized as follows.
In Sec. II, we introduce our model and show relevant formulas for phenomenology.
In Sec. III, we show our phenomenological analysis of muon $g-2$, LFVs and collider physics.
 Finally we devote the summary of our results and the conclusion.

\begin{table}[t!]
\begin{tabular}{|c||c|c|c|c|c|c|c|c|c||c|c|c|}\hline\hline  
& ~$Q_L^a$~& ~$u_R^a$~  & ~$d_R^a$~& ~$L_L^a$~& ~$e_R^a$~ & ~$N_R^a$~ & ~$N_L^a$~ 
& ~$E_L^a$~ & ~$E_R^a$~ & ~$H$~ & ~$\varphi_1$~ & ~$\varphi_2$~ \\\hline\hline 
$SU(3)_C$ & $\bm{3}$  & $\bm{3}$ & $\bm{3}$ & $\bm{1}$ & $\bm{1}$ & $\bm{1}$ & $\bm{1}$ & $\bm{1}$ & $\bm{1}$ & $\bm{1}$ & $\bm{1}$  & $\bm{1}$  \\\hline 
$SU(2)_L$ & $\bm{2}$  & $\bm{1}$  & $\bm{1}$  & $\bm{2}$  & $\bm{1}$ & $\bm{1}$  & $\bm{1}$ & $\bm{1}$  & $\bm{1}$  & $\bm{2}$ & $\bm{1}$  & $\bm{1}$   \\\hline 
$U(1)_Y$   & $\frac16$ & $\frac23$ & $-\frac13$ & $-\frac12$  & $-1$ & $0$ & $0$ & $-1$ & $-1$  & $\frac12$  & $0$ & $0$\\\hline
$U(1)_{R}$   & $0$ & $1$ & $-1$   & $0$  & $0$  & $1$  & $0$ & $0$  & $-1$  & $1$  & $-1$ & $-\frac12$\\\hline
\end{tabular}
\caption{ 
Charge assignments of the our fields
under $SU(3)_C\times SU(2)_L\times U(1)_Y\times U(1)_{R}$, where its upper index $a$ is the number of family that runs over $1-3$.}
\label{tab:1}
\end{table}

\section{Model setup and Constraints}
Here, we review our model.
We introduce three vector-like neutral fermions $N_{L,R}$ and singly-charged fermions $E_{L,R}$,
where only $N_R$ and $E_R$ has nonzero $U(1)_R$ charge with $1$ and $-1$, respectively.
It suggests that masses of charged-lepton arise not via the SM type but via mixings among singly-charged fermions. 
One can straightforwardly confirm the typical four patterns of  chiral anomaly cancellations per one generation; $[U(1)_R]= [U(1)_R]^3= [U(1)_R]^2[U(1)_Y]= [U(1)_R][U(1)_Y]^2=0$.~\footnote{In our paper, quark sector is the same as the SM except for the fact that right-handed ones interact with $Z'$ boson.}
As for scalar sector, we introduce an isospin singlet fields $\varphi_1$ and $\varphi_2$ with $U(1)_R$ charge $-1$ and $-\frac12$ to spontaneous $U(1)_R$  symmetry breaking
where SM-like Higgs $H$ also has nonzero charge under  $U(1)_R$ in order to induce Yukawa Lagrangian. Each of VEVs is denoted by $\langle H\rangle\equiv [0,v/\sqrt2]^T$ and  $\langle \varphi_{1(2)} \rangle\equiv v'_{1(2)}/\sqrt2$.  
All the field contents and their assignments are summarized in Table~\ref{tab:1}.

\subsection{Lagrangian and scalar masses \label{sec:yukawa}}
The relevant lepton Yukawa Lagrangian under these symmetries is given by 
\begin{align}
&-{\cal L}_{Y}
= 
 (y_\ell)_{ab} \bar L^a_L H E^b_R
+ (y_E)_{aa} \varphi_1 \bar E^a_L E^a_R
+ (m_{Ee})_{ab} \bar E^a_L e^b_R
 \nn\\
& \qquad \quad +  (y_D)_{ab} \bar L^a_L\tilde H N^b_R  
+  (y_{N})_{aa} \varphi_1 \bar N^{a}_L N^a_R
+  (M_{N_L})_{ab} \bar N^{a}_L N^{c b}_L
+ {\rm h.c.}, \label{Eq:yuk} 
\end{align}
where $\tilde H\equiv i\sigma_2H$, and upper(lower) indices $(a,b)=1$-$3$ for fields(Yukawa or mass matrix) are the number of families, and $y_E,\ y_N$ can be diagonal matrix without loss of generality.

The scalar potential is given by
\begin{align}
 {\cal V} = & \mu^2_1 |\varphi_1|^2 - \mu^2_2 |\varphi_2|^2 -\mu^2_H |H|^2 + \lambda_H |H|^4 
 + \lambda_1 |\varphi_1|^4 + \lambda_2 |\varphi_2|^4 - \mu_3 ( \varphi_1^* \varphi_2 \varphi_2  + h.c.)  \nonumber \\
&  + \lambda_{3} |\varphi_1|^2 |H|^2 +  \lambda_{4} |\varphi_2|^2 |H|^2 + \lambda_{5} |\varphi_1|^2  |\varphi_2|^2.\label{Eq:pot}
\end{align}
The scalar fields are parameterized as 
\begin{align}
&H =\left[\begin{array}{c}
w^+\\
\frac{v + r +i z}{\sqrt2}
\end{array}\right],\quad 
\varphi_{1,2}=
\frac{v'_{1,2}+ r'_{1,2} + iz'_{1,2} }{\sqrt2},
\label{component}
\end{align}
where $w^+$ and $z$ are massless Nambu-Goldstone(NG) bosons which are absorbed by the SM gauge bosons $W^+$ and $Z$, and
one linear combination of $z'_{1,2}$ corresponds to NG boson abosorbed by an extra $Z'$ boson from $U(1)_R$.  
The VEVs are obtained from tadpole conditions $\frac{\partial {\cal V}}{\partial v} = \frac{\partial {\cal V}}{\partial v'_1} =\frac{\partial {\cal V}}{\partial v'_2} =0$. 
We obtain condition for $v'_1$ from $\frac{\partial {\cal V}}{\partial v'_1} =0$ such that 
\begin{equation}
\mu_1^2 v'_1 + \lambda_1 v'^3_1 - \frac{1}{2 \sqrt2} \mu_3 v'^2_2 + \lambda_5 v'_1 v'^2_2 = 0.
\end{equation}
Here the VEV of $\varphi_1$ is approximately given by
\begin{equation}
v'_1 \simeq \frac{1}{2 \sqrt2} \frac{\mu_3 v'^2_2}{ \mu_1^2 + \lambda_5 v'^2_2},
\end{equation}
where we assume $v'_1$ is much smaller than $v'_2$ and $v'^3_1$ term is ignored.
The hierarchy of VEVs is consistently achieved by choosing $\mu_3 v'_2 \ll \mu_1^2$ and/or $\mu_3 v'_2 \ll \lambda_5 v'^2_2$.
This VEV hierarchy is necessary to explain W boson mass anomaly and to obtain sizable muon $g-2$ at the same time, as we discuss below.
In this case, $z'_2$ corresponds to the NG boson to be absorbed by $Z'$ boson.

In our analysis, we assume $\lambda_4$ and $\lambda_5$ to be negligibly small for simplicity, and only $r$ and $r'_1$ mixes.
Then, we obtain the mass matrix for CP even scalar, $m_R^2$, in the basis of $(r,r'_1)$, where 
the mass eigenstates $\{ h,H \}$ is found to be $(r,r'_1)^T=O_R (h, H)^T$,
and mass eigenvalues are given by $m_{h,H}^2=O_R^T m_R^2 O_R$.
$m_R^2$ and $O_R$ are obtained as
\begin{align}
m_{R}^2
& \simeq
\left[\begin{array}{cc}
2 v^2\lambda_{H} &  v v'_1 \lambda_{3}  \\ 
v v'_1 \lambda_{3} & \mu_1^2 \\ 
\end{array}\right],
O_R
=
\left[\begin{array}{cc}
c_\theta &  s_\theta  \\ 
-s_\theta &  c_\theta \\ 
\end{array}\right],
\end{align}
where $c_\theta(s_\theta)$ stands for $\cos \theta (\sin \theta)$ with $s_{2\theta}=\frac{2vv'_1 \lambda_3}{m^2_{h}-m^2_{H}}$. 
The mass eigenvalues are also calculated such that
\begin{equation}
m_{h,H}^2 \simeq (v^2 \lambda_H + \mu_1^2) \mp \sqrt{(v^2 \lambda_H - \mu_1^2)^2 + v^2 v'^2_1 \lambda_3^2}.
\end{equation}
Here $h\equiv h_{SM}$ is the SM Higgs, therefore,  $m_{h}=$125 GeV.
The mixing effect for CP-even scalar is constrained by the measurements of Higgs production cross section and its decay branching ratio at the LHC, 
and $s_a\lesssim 0.3$ is provided by the current data~\cite{pdg}.
The mass of $r'_2$ is approximately given by $m_{r'_2} \simeq \sqrt{2 \lambda_2} v'_2$ which is supposed to be much heavier than $m_H$.

\subsection{Oblique parameters \label{sec:oblique}}
Oblique parameters come from {\it $Z_{SM}-Z'$ mixing}.
Thus, we firstly discuss this effect.
Since $H$ has nonzero $U(1)_R$ charge, there is mixing between $Z_{SM}$ and $Z'$. 
The resulting mass matrix in basis of $(Z_{SM},Z')$  is given by
\begin{align}
m_{Z_{SM}Z'}^2
&\simeq \frac{1}4
\left[\begin{array}{cc}
(g_1^2+g_2^2) v^2 &  -2\sqrt{g_1^2+g_2^2}g' v^2  \\ 
-2 \sqrt{g_1^2+g_2^2}g' v^2  & 4 g'^2 (v^2+  v'^2_2)   \\ 
\end{array}\right]
=
m_{Z'}^2
\left[\begin{array}{cc}
\epsilon_1^2 & -\epsilon_1 \epsilon_2  \\ 
-\epsilon_1 \epsilon_2 & 1+\epsilon_2^2  \\ 
\end{array}\right],
\end{align}  
where $m_{Z_{SM}}\equiv \frac{\sqrt{g_1^2+g_2^2}v}{2}$, $m_{Z'}\equiv  g'v'_2$, $\epsilon_1\equiv \frac{m_{Z_{SM}}}{m_{Z'}}$, $\epsilon_2\equiv \frac{v}{v'_2}$, $g_1$, $g_2$, and $g'$ are gauge coupling of $U(1)_Y$, $SU(2)_L$, and $U(1)_R$, respectively.
Note that we ignored $v'_1$ in $m_{Z'}$ formula due to the relation $v'_1 \ll v'_2$.
Then its mass matrix is diagonalized by the two by two mixing matrix $V$ as $V m_{Z_{SM}Z'}^2 V^T
\equiv {\rm Diag}(m^2_{Z_{}},m^2_{Z_{R}}) $,
where we work under $\epsilon_2^2 \ll 1$ and
\begin{align}
m^2_{Z}&\approx m_{Z_{SM}}^2(1-\epsilon_2^2),\
m^2_{Z_{R}}\approx m_{Z'}^2 (1 +  \epsilon_1^2\epsilon_2^2),\label{eq:zm}
\\
V&\approx
\left[\begin{array}{cc}
c_{Z} &  s_{Z} \\ 
-s_{Z}  &  c_{Z}  \\ 
\end{array}\right], \quad \theta_{Z} = \frac{1}{2} \tan^{-1} \left[ \frac{2 \epsilon_1 \epsilon_2}{1+\epsilon_2^2-\epsilon_1^2} \right].
\end{align} 
{The $Z_R$ mass can be approximated as $m_{Z_R} \simeq m_{Z'} =  g' v'_2$ since $\epsilon_{1,2}$ is small. Thus gauge coupling $g'$ is almost fixed if we choose values of $m_{Z_R}$ and $v'$. }

%
 In our case,
only $\Delta T$ is nonzero induced via $Z-Z'$ mixing thanks to zero $U(1)_R$ charges of $L_L,\ e_R$ and defined by
\begin{align}
\Delta T = \frac{1}{\alpha_{\rm em}}
 \frac{m_{Z_{SM}}^2-m_{Z}^2}{m_{Z_{SM}}^2}
 \simeq \frac{\epsilon^2_2}{\alpha_{\rm em}},
  \label{eq:ST}%
\end{align}
where we have used Eq.~(\ref{eq:zm}) in the last part of the above equation.
Thus, $\Delta T$ is straightforwardly given by inserting $\epsilon_2\equiv v/v'_2$,
$ v'_2 = m_{Z'}/g'$,  $v=  2 m_Z \cos \theta_W/g_2$ with $\theta_W$ being the Weinberg angle
and simply given by~\cite{Strumia:2022qkt}~\footnote{Notice here that the other valid oblique parameters $\Delta S,\ \Delta W,\ \Delta Y$ are zero.}
\begin{align}
\Delta T \simeq \frac{v^2}{\alpha_{\rm em}}
 \frac{ g'^2}{m_{Z'}^2} = \frac{4 m_Z^2 \cos^2 \theta_W}{g_2^2 \alpha_{\rm em}}
 \frac{ g'^2}{m_{Z'}^2}.  \label{eq:ST}
\end{align}
Note also that contribution to $\Delta T$ at loop level is negligible since our new particles are $SU(2)$ singlet. Although mixing between the SM charged lepton and new charged lepton discussed below can affect the $\Delta T$ at loop level it is highly suppressed by small mixing angle. 
Thus our relevant parameters for $\Delta T$ are only new gauge coupling $g'$ and $Z'$ mass $m_{Z'}$.
From global fit including CDF II W boson mass with $S \sim 0$ we obtain $1 \sigma$ range of $T$ as 
\begin{equation}
0.09 \leq \Delta T \leq 0.14.
\label{eq:delT}
\end{equation}
Then it is found from Eq.~\eqref{eq:ST}:
\begin{equation}
\label{eq:range-vphi}
20 \ {\rm TeV} \lesssim \frac{m_{Z'}}{g'} (= v'_2) \lesssim 31 \ {\rm TeV},
\end{equation}
where we used central value of Eq.~\eqref{eq:Wmass} for $m_W$.
This range is allowed by LEP constraints~\cite{ALEPH:2013dgf} and dijet searches at the LHC~\cite{Dobrescu:2021vak,ATLAS:2019fgd,CMS:2019gwf}.
Here, we emphasize that our model only provides a sizable contribution to $\Delta T$ via $Z$--$Z'$ mixing. Thus modification of $W$ boson mass is characterized by $\Delta T$ that is estimated by $Z'$ mass, new gauge coupling and other electroweak observables $\{m_Z, \theta_W, \alpha_{\rm em} \}$. It is the advantage of our model that other oblique parameters are not modified.

Here we comment on the situation in the case of original $U(1)_R$ model.
In this case, we have contributions to other oblique parameters $W$ and $Y$ due to $Z'$ interactions with charged leptons~\cite{Cacciapaglia:2006pk}. 
The oblique parameters, including $W$ and $Y$ are defined from effective Lagrangian 
\begin{equation}
\mathcal{L} = -\frac12 W^3_\mu \Pi_{33}(p^2) W^{3 \mu} - \frac12 B_\mu \Pi_{00} (p^2) B^\mu - W^3_\mu \Pi_{30} (p^2) B^\mu - W^+_\mu \Pi_{WW}(p^2) W^{- \mu},
\end{equation}
where $W^3_\mu$, $B^\mu$ and $W^\pm$ are gauge fields corresponding to the third component of $SU(2)_L$, $U(1)_Y$  and $W$ boson respectively, 
and $p^2$ is the momentum square carried by the gauge field.
The oblique parameters $W$ and $Y$ are then given by
\begin{equation}
W = \frac{m^2_W}{2} \frac{d^2}{d(p^2)^2} \Pi_{33}|_{p^2 =0}, \quad Y = \frac{m^2_W}{2} \frac{d^2}{d(p^2)^2} \Pi_{00}|_{p^2 =0}.
\end{equation}
It is shown in ref.~\cite{Cacciapaglia:2006pk} that $W=Y=0$ is obtained when leptons are not charged under new $U(1)$ symmetry, which is our case.
Otherwise, the values of $W$ and $Y$ have similar order as $\hat T \equiv \alpha_{\rm em} T$ and the 
LHC constraints are $|W| \lesssim 1.8 \times 10^{-4}$ and $|Y| \lesssim 2.0 \times 10^{-4}$ at $1 \sigma$ confidence level~\cite{ATLAS:2017fih, CMS:2022yjm, Strumia:2022qkt}.
Thus the parameter region realizing $\Delta T$ in Eq.~\eqref{eq:delT} is excluded by the constraints.
Therefore we need leptophobic $U(1)_R$ to explain the W boson anomaly.

\subsection{Charged-lepton sector}
The charged-lepton mass matrix is obtained via mixing among the singly-charged fermions, and 
the form is give by~\cite{Dey:2019cts}
\begin{align}
& \left( \begin{array}{c} \bar e_L \\ \bar E_L \end{array} \right)^T
{\cal M_{E}}
\left( \begin{array}{c}  e_R \\  E_R \end{array} \right)
=
\left( \begin{array}{c} \bar e_L \\ \bar E_L \end{array} \right)^T
\left[\begin{array}{cc}
0 & m_{eE}  \\ 
m_{Ee}  & M_E  \\ 
\end{array}\right]
\left( \begin{array}{c}  e_R \\  E_R \end{array} \right),\\
& {\cal M_{E}}{\cal M_{E}}^\dag
=
\left[\begin{array}{cc}
 m_{eE} m_{eE}^\dag & m_{eE} M_E  \\ 
M_E m_{eE}^\dag & M_{E}^2+m_{Ee} M_{Ee}^\dag  \\ 
\end{array}\right], \quad 
{\cal M_{E}}^\dag {\cal M_{E}}
=
\left[\begin{array}{cc}
 m_{Ee}^\dag m_{Ee} & m_{Ee}^\dag M_E  \\ 
M_E m_{Ee} & M_{E}^2+ m_{eE}^\dag m_{eE} \\ 
\end{array}\right],
\end{align}
where $m_{eE}\equiv y_\ell v/\sqrt2$, $M_{E}\equiv y_E v'/\sqrt2$.
The mass matrix is diagonalized by the transformation $(e_{L(R)}, E_{L(R)}) \to V_{L(R)}\ell_{L(R)}$.
Thus we can obtain diagonalization matrices $V_{L}$ and $V_{R}$ which respectively diagonalize $M_{E} M_{E}^\dag$ and $M_{E}^\dag M_{E}$ as $V_{L}^\dag {\cal M_{E}} {\cal M_{E}}^\dag V_L =V_{R}^\dag {\cal M_{E}}^\dag {\cal M_{E}} V_{R}={\rm diag} |D_{E_a}|^2$(a=1-6), where the first three mass eigenstates correspond to the SM charged-leptons.
We then write ${\rm diag} D_{E_a} = {\rm diag}(m_e, m_\mu, m_\tau, M_{E_1},  M_{E_2},  M_{E_3})$.

\subsection{New contribution to $Z\to \ell\bar\ell$}
Due to the mixing between the exotic singly-charged fermions and the SM leptons, we have new contribution to $Z\to \ell_i\bar\ell_j$.
Their kinetic Lagrangian in terms of mass eigenvectors is given by 
\begin{align}
\frac{g_2}{c_W}
\left[\left(-\frac12 \sum_{a=1}^3V^\dag_{L_{ia}} V_{L_{aj}} +s_W^2\delta_{ij}\right) \bar \ell_{L_i} \gamma^\mu \ell_{L_j} 
+s_W^2 \delta_{ij} \bar \ell_{R_i} \gamma^\mu \ell_{R_j}
\right] Z_\mu,
\end{align}
where $s_W,\ c_W$ are short-hand notations of Weinberg angles; $\sin\theta_W,\ \cos\theta_W$, which are rewritten in terms of $g_1,\ g_2$. 
Then, the decay rate of $Z\to \ell_i\bar\ell_j$ is given by
\begin{align}
\Gamma(Z\to \ell_i\bar\ell_j)
\simeq
\frac{g_2^2}{24\pi c_W^2}m_Z
\left[
\left|-\frac12 \sum_{a=1}^3V^\dag_{L_{ia}} V_{L_{aj}} +s_W^2\delta_{ij}\right|^2 +s_W^4 \delta_{ij}
\right],
\end{align}
where we assume $m_Z>>m_\ell$.
Here, one confirms that the SM contribution is derived when $ \sum_{a=1}^3V^\dag_{L_{ia}} V_{L_{aj}}=1$;
\begin{align}
\Gamma(Z\to \ell_i\bar\ell_j)_{SM}
\simeq
\frac{g_2^2}{24\pi c_W^2}m_Z
\left[ \frac14 -s_W^2  + 2 s_W^4 \right]\delta_{ij}. 
\end{align}
Thus, the new contribution of branching ratios are given by
\begin{align}
\Delta {\rm BR}(Z\to \ell_i\bar\ell_j)=
\frac{\Gamma(Z\to \ell_i\bar\ell_j)-
\Gamma(Z\to \ell_i\bar\ell_j)_{SM}}
{\Gamma_{Z}^{\rm tot}}\quad {\rm for \ i= j},\\
{\rm BR}(Z\to \ell_i\bar\ell_j)=
\frac{\Gamma(Z\to \ell_i\bar\ell_j)} {\Gamma_{Z}^{\rm tot}}\quad {\rm for \ i\neq j},
\end{align}
where the total $Z$ decay width $\Gamma_{Z}^{\rm tot} = 2.4952 \pm 0.0023$~GeV~\cite{pdg}.

The current bounds on the lepton-flavor-(conserving)changing $Z$ boson decay 
branching ratios(BRs) at 95 \% CL are given by \cite{pdg}:
\begin{align}
 & \Delta {\rm BR}(Z\to e^\pm e^\mp) < \pm4.2\times10^{-5} ~,\
 \Delta   {\rm BR}(Z\to \mu^\pm\mu^\mp) <  \pm6.6\times10^{-5} ~,\nn\\
&   \Delta   {\rm BR}(Z\to \tau^\pm\tau^\mp) <  \pm8.3\times10^{-5} ~,\
    {\rm BR}(Z\to e^\pm\mu^\mp) < 7.5\times10^{-7} ~,\nn\\
&  {\rm BR}(Z\to e^\pm\tau^\mp) < 9.8\times10^{-6} ~, \ 
  {\rm BR}(Z\to \mu^\pm\tau^\mp) < 1.2\times10^{-5} ~.\label{eq:zmt-cha}
\end{align}

\begin{figure}[tb]
\begin{center}
\includegraphics[width=97.0mm]{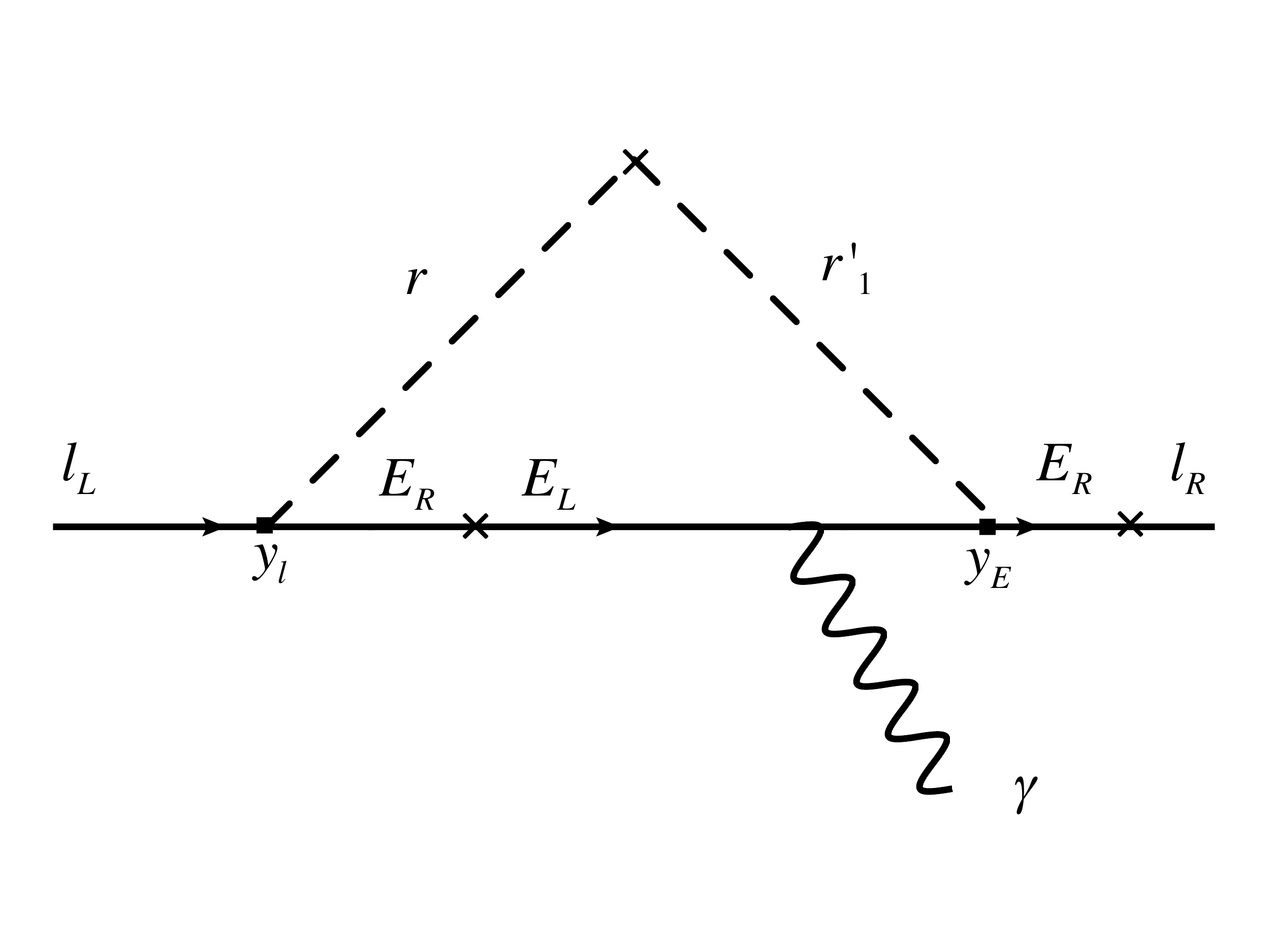} 
\caption{The diagram that gives dominant contributions to $\ell \to \ell' \gamma$ and muon $g-2$ in flavor basis. The $\times$ mark indicates mass (mixing) insertion.}
  \label{fig:diagram}
\end{center}\end{figure}

\subsection{Lepton flavor violations and muon $g-2$}
Due to mixing between the SM charged-lepton and heavier leptons, we have nonzero LFVs and muon $g-2$ via $y_\ell$.
The current experimental upper bounds on LFVs are given 
by~\cite{MEG:2016leq,MEG:2013oxv}
  \begin{align}
  {\rm BR}(\mu\rightarrow e\gamma) &\leq4.2\times10^{-13},\quad 
  {\rm BR}(\tau\rightarrow \mu\gamma)\leq4.4\times10^{-8}, \quad  
  {\rm BR}(\tau\rightarrow e\gamma) \leq3.3\times10^{-8}~.
 \label{expLFV}
 \end{align}
On the other hand, new results on the muon $(g-2)$ were recently published by the E989 collaboration at Fermilab \cite{Muong-2:2021ojo}: 
\begin{align}
a^{\rm FNAL}_\mu =116592040(54) \times 10^{-11}.
\label{exp_dmu}
\end{align}
Combined with the previous BNL result, this means that the muon $(g-2)$ deviates from the SM prediction by 4.2$\sigma$ level~\cite{Muong-2:2021ojo, Aoyama:2012wk,Aoyama:2019ryr,Czarnecki:2002nt,Gnendiger:2013pva,Davier:2017zfy,Keshavarzi:2018mgv,Colangelo:2018mtw,Hoferichter:2019mqg,Davier:2019can,Keshavarzi:2019abf,Kurz:2014wya,Melnikov:2003xd,Masjuan:2017tvw,Colangelo:2017fiz,Hoferichter:2018kwz,Gerardin:2019vio,Bijnens:2019ghy,Colangelo:2019uex,Blum:2019ugy,Colangelo:2014qya,Hagiwara:2011af}:
\begin{align}
\Delta a^{\rm new}_\mu = (25.1\pm 5.9)\times 10^{-10} ,
\label{exp_dmu}
\end{align}
and it could be a verfiable signature of the physics beyond the SM.

In our scenario, the relevant Lagrangian to induce LFVs and muon $g-2$ in terms of mass eigenstate is given by~\footnote{Even though we have a contribution from a new gauge boson $Z'$ to these phenomenologies, these are subdominant for the value of $g'/m_{Z'}$ in the range of Eq.~\eqref{eq:range-vphi}. We have checked it numerically.}
\begin{align}
& \mathcal{L}_{LFV}
=
(Y_{\alpha\beta} c_\theta -\tilde Y_{\alpha\beta}  s_\theta)h \bar \ell_\alpha P_R \ell_\beta
+
(Y_{\alpha\beta} s_\theta +\tilde Y_{\alpha\beta}  c_\theta)H \bar \ell_\alpha P_R \ell_\beta + h.c., \\
& Y_{\alpha \beta} \equiv \frac1{\sqrt2} \sum_{a=1,2,3} \sum_{b=1,2,3} (V^\dag_{L})_{\alpha,a} (y_\ell)_{ab}(V_R)_{b+3,\beta}, \\
& \tilde{Y}_{\alpha\beta} \equiv  \frac1{\sqrt2} \sum_{a=1,2,3} \sum_{b=1,2,3}  (V^\dag_{L})_{\alpha,a+3} (y_E)_{aa}(V_R)_{a+3,\beta}.
%
\end{align}
Then, the dominant contributions to LFVs and muon $g-2$ at one-loop level are given by 
\begin{align}
& {\rm BR}(\ell_\beta\to\ell_\alpha\gamma)
\simeq
\frac{12\pi^2 C_{\beta\alpha}}{(4\pi)^4 m_{\ell_\beta}^2 G_F^2}
(|a_{L_{\alpha\beta}}|^2+|a_{R_{\alpha\beta}}|^2),\\
& \Delta a_\mu\simeq
-\frac{m_\mu}{(4\pi)^2}(a_{L_{22}}+a_{R_{22}}),\\
&a_{L_{\alpha\beta}}\approx 
-\frac14
{(Y^\dag_{\alpha\rho}c_\theta -\tilde Y^\dag_{\alpha\rho}s_\theta) D_{E_\rho} 
(Y^\dag_{\rho \beta}c_\theta -\tilde Y^\dag_{\rho \beta}s_\theta)} F(h_1,D_{E_\rho})
\nn\\&
-\frac14
{(Y^\dag_{\alpha\rho}s_\theta +\tilde Y^\dag_{\alpha\rho}c_\theta) D_{E_\rho} 
(Y^\dag_{\rho \beta}s_\theta +\tilde Y^\dag_{\rho \beta}c_\theta)} F(h_2,D_{E_\rho}),\\
%
%
&a_{R_{\alpha\beta}}\approx 
-\frac14
{(Y_{\alpha\rho}c_\theta -\tilde Y_{\alpha\rho}s_\theta) D_{E_\rho} 
(Y_{\rho \beta}c_\theta -\tilde Y_{\rho \beta}s_\theta)} F(h_1,D_{E_\rho})
\nn\\&
-\frac14
{(Y_{\alpha\rho}s_\theta +\tilde Y_{\alpha\rho}c_\theta) D_{E_\rho} 
(Y_{\rho \beta}s_\theta +\tilde Y_{\rho \beta}c_\theta)} F(h_2,D_{E_\rho}),\\
&
F(m_{h_i},D_{E_\rho}) \approx
\frac{m_{h_i}^4 -4m_{h_i}^2 D_{E_\rho}^2 + 3D_{E_\rho}^4-2m_{h_i}^2(m_{h_i}^2-2 D_{E_\rho}^2)\ln\left[\frac{m_{h_i}^2}{D_{E_\rho}^2}\right] }{(m_{h_i}^2- D_{E_\rho}^2)^3},
\end{align}
where $C_{21}\approx1$, $C_{31}\approx0.1784$, $C_{32}\approx0.1736$, 
$h_1\equiv h$, $h_2\equiv H$.
The dominant contribution is obtained from the diagram in Fig.~\ref{fig:diagram} that does not have chiral suppression.

\subsection{Neutrino mass via inverse seesaw mechanism}

After spontaneous gauge symmetry breaking, we obtain neutral fermion mass matrix in the basis of $(\nu^c_L, N_R, N^c_L)$ as follows 
\begin{equation}
M_N = \left[ \begin{array}{ccc} 0 & m_D & 0 \\ m^T_D & 0 & M \\ 0 & M & M_{N_L} \end{array} \right],
\end{equation}
where $m_D \equiv v y_D/\sqrt2$ and $M \equiv y_N v'_1/\sqrt2$.
When mass parameters satisfy $M_{N_L} \ll  m_D \lesssim M$ active neutrino mass can be approximately written by
\begin{equation}
m_\nu \simeq m_D M^{-1} M_{N_L} (M^T)^{-1} m^T_D.
\end{equation}
The neutrino mass matrix is diagonalized by a unitary matrix $U_\nu$; $D_\nu = U^T_\nu m_\nu U_\nu$ with $D_\nu \equiv {\rm diag}(m_1,m_2,m_3)$.
Including charged lepton mixing matrix, the PMNS matrix is defined by $U_{PMNS} \equiv V^\dagger_L U_\nu$. 

We discuss constraint from non-unitarity which is described by a matrix $U'_{PMNS}$.
This matrix is typically parametrized by the form of 
\begin{align}
U'_{PMNS}\equiv \left(1-\frac12 FF^\dag\right) U_{PMNS},
\end{align}
where $F\equiv  (M^{T})^{-1} m_D$ is a hermitian matrix. 
The global constraints on elements of $|FF^\dagger|$ are found combining several experimental results such as the SM $W$ boson mass $M_W$, the effective Weinberg angle $\theta_W$, several ratios of $Z$ boson fermionic decays, invisible decay of $Z$, electroweak universality, measured Cabbibo-Kobayashi-Maskawa, and lepton flavor violations~\cite{Fernandez-Martinez:2016lgt}.
The result is then given by~\cite{Agostinho:2017wfs}
\begin{align}
|FF^\dagger|\leq  
\left[\begin{array}{ccc} 
2.5\times 10^{-3} & 2.4\times 10^{-5}  & 2.7\times 10^{-3}  \\
2.4\times 10^{-5}  & 4.0\times 10^{-4}  & 1.2\times 10^{-3}  \\
2.7\times 10^{-3}  & 1.2\times 10^{-3}  & 5.6\times 10^{-3} \\
 \end{array}\right].
\end{align} 
If we require $|F| \sim 10^{-5}$ conservatively, $M_{N_{L}} \sim 1-10$ GeV can reproduce active neutrino mass scale.
Also we require $y_D \sim 10^{-4}$ for $M = \mathcal{O}(1)$ TeV.
Observed neutrino mixing can be easily obtained since we have sufficient number of free parameters $\{y_D, y_N, M_{N_L} \}$.

\section{Numerical analysis and phenomenological consequences}
 
 In this section we carry out numerical study to estimate muon $g-2$ and LFVs. 
 We also calculate $Z'$ production cross section for parameter space explaining W boson mass anomaly and 
 show phenomenological implications of our scenario at hadron collider experiments.
 
 \subsection{Parameter scan for muon $g-2$ and LFVs}

Here we discuss muon $g-2$ taking into account charged lepton masses and LFV constraints in our model performing numerical analysis.
The dominant contribution to muon $g-2$ comes from the diagram in Fig.~\ref{fig:diagram} since it has chirality change inside loop 
giving heavy charged lepton mass factor.
The relevant free parameters are 
\begin{equation}
\{ (y_{\ell})_{ab},  (y_E)_{aa}, (m_{Ee})_{ab},  \sin \theta, m_H \},
\end{equation}
where $a,b = 1-3$.
We then scan these free parameters globally to search for best fit value of muon $g-2$.
In Table~\ref{tab:BP}, we show our benchmark point(BP) found by numerical analysis.
We find that large value of Yukawa coupling $y_E$ is preferred to obtain sizable muon $g-2$ 
while exotic heavy charged lepton masses are around 700 GeV.
Then values of $v'_1$ and $m_H$ are preferred to be around electroweak scale.
This is the reason why we need two singlet scalars $\varphi_1$ and $\varphi_2$ 
to explain both muon $g-2$ and W boson mass anomaly as the required scale of 
these VEVs are different; $v'_1 \ll v'_2$.
 We also show 
 ${\rm BR}(\ell \to \ell' \gamma)$
 and find that those of $\mu \to e\gamma$ and $\tau \to \mu \gamma$ are close to the current upper limit.
 Thus it could be tested in future measurements.
 Furthermore we show $(\Delta) {\rm BR}(Z \to \ell \ell')$ values that might be also tested in future precision measurements 
 for $Z$ boson decay.
 
 \begin{center} 
\begin{table}[tb]
\begin{tabular}{|c|c|}\hline 
\multicolumn{2}{|c|}{Input} \\ \hline
$v'_1/$GeV & $284$  \\ \hline
$m_H/$GeV & $245$  \\ \hline
$\sin \theta$ & 0.250 \\ \hline
$[(y_\ell)_{11}, (y_\ell)_{12},  (y_\ell)_{13}]$ & $[-0.000512, \  0.00520, \ 0.00236]$ \\ \hline
$[(y_\ell)_{21}, (y_\ell)_{22},  (y_\ell)_{23}]$ & $[-0.000105, \  0.000624, \ 0.0643]$ \\ \hline
$[(y_\ell)_{31}, (y_\ell)_{32},  (y_\ell)_{33}]$ & $[0.000148, \  0.0145, \ 0.0647]$ \\ \hline
$[(m_{eE})_{11}, (m_{eE})_{12}, (m_{eE})_{13}]/$GeV & $[0.674, \ 16.8, \ 5.92]$ \\ \hline
$[(m_{eE})_{21}, (m_{eE})_{22}, (m_{eE})_{23}]/$GeV & $[-16.2, \ -46.3, \ 22.3]$ \\ \hline
$[(m_{eE})_{31}, (m_{eE})_{32}, (m_{eE})_{33}]/$GeV & $[-12.2, \ 69.5, \ 19.8]$ \\ \hline
$[(y_E)_{11}, (y_E)_{22},  (y_E)_{33}]$ & $[-3.26, \  3.43, \ -3.41]$ \\ \hline \hline
\multicolumn{2}{|c|}{Output} \\ \hline
$[m_{E_1}, m_{E_2}, m_{E_3}]/$GeV & $[655, \ 690, \ 694]$ \\ \hline
$\Delta a_\mu$ & $2.11 \times 10^{-9}$ \\ \hline
${\rm BR}(\mu \to e \gamma)$ & $3.59 \times 10^{-13}$ \\ \hline
${\rm BR}(\tau \to e \gamma)$ & $9.91 \times 10^{-11}$ \\ \hline
${\rm BR}(\tau \to \mu \gamma)$ & $2.22 \times 10^{-8}$ \\ \hline
$\Delta {\rm BR}(Z \to e^\pm e^\mp)$ & $-1.55 \times 10^{-9}$ \\ \hline
$\Delta {\rm BR}(Z \to \mu^\pm \mu^\mp)$ & $-6.11 \times 10^{-7}$ \\ \hline
$\Delta {\rm BR}(Z \to \tau^\pm \tau^\mp)$ & $-3.82 \times 10^{-5}$ \\ \hline
$\Delta {\rm BR}(Z \to e^\pm \mu^\mp)$ & $6.27 \times 10^{-17}$ \\ \hline
$\Delta {\rm BR}(Z \to e^\pm \tau^\mp)$ & $3.83 \times 10^{-14}$ \\ \hline
$\Delta {\rm BR}(Z \to \mu^\pm \tau^\mp)$ & $3.64 \times 10^{-11}$ \\ \hline
\end{tabular}
\caption{Benchmark point that explains muon $g-2$.}
\label{tab:BP}
\end{table}
\end{center}

\begin{figure}[tb]
\begin{center}
\includegraphics[width=97.0mm]{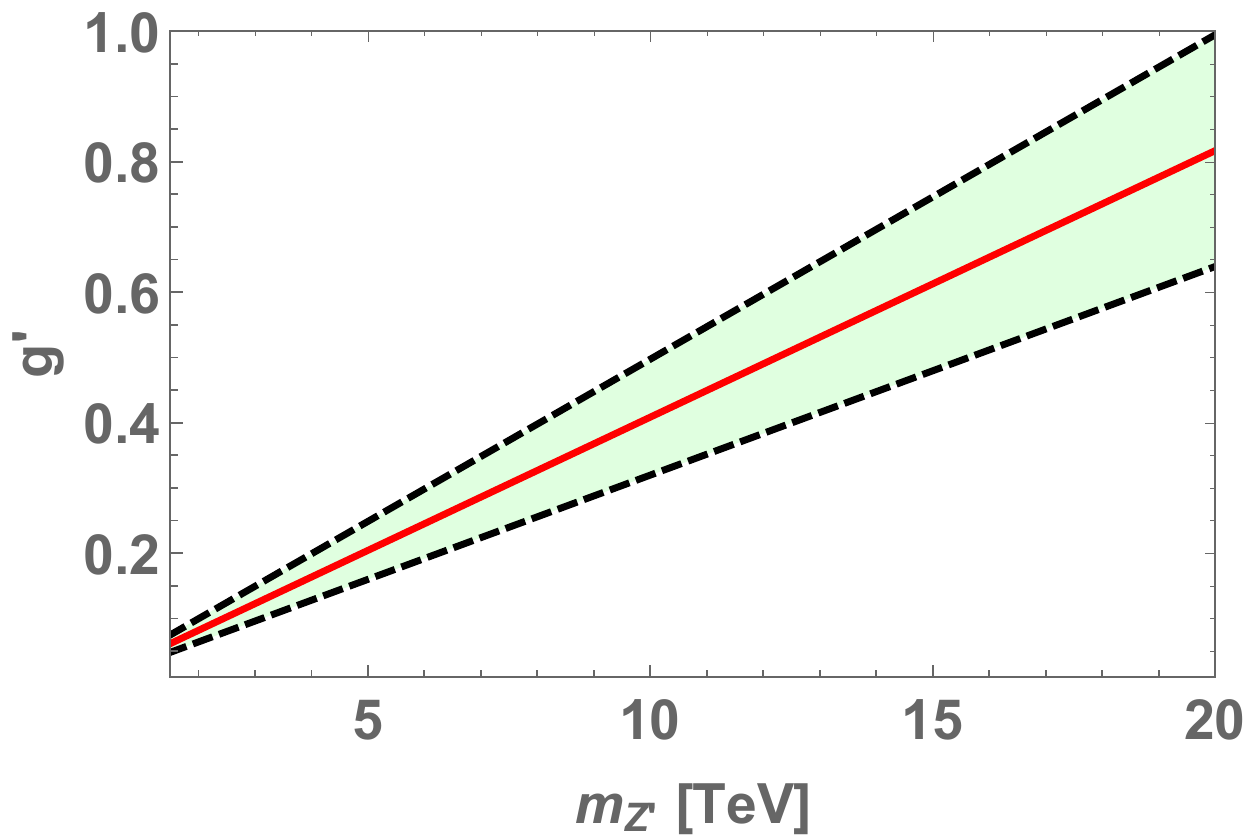} 
\caption{The $1 \sigma$ region to explain W boson mass anomaly on $\{m_{Z'}, g'\}$ plane where the red colored 
 line indicates the value providing central value of $\Delta T$ in Eq.~\eqref{eq:delT}. }
  \label{fig:gx-mzp}
\end{center}\end{figure}
 
\begin{figure}[tb]
\begin{center}
\includegraphics[width=97.0mm]{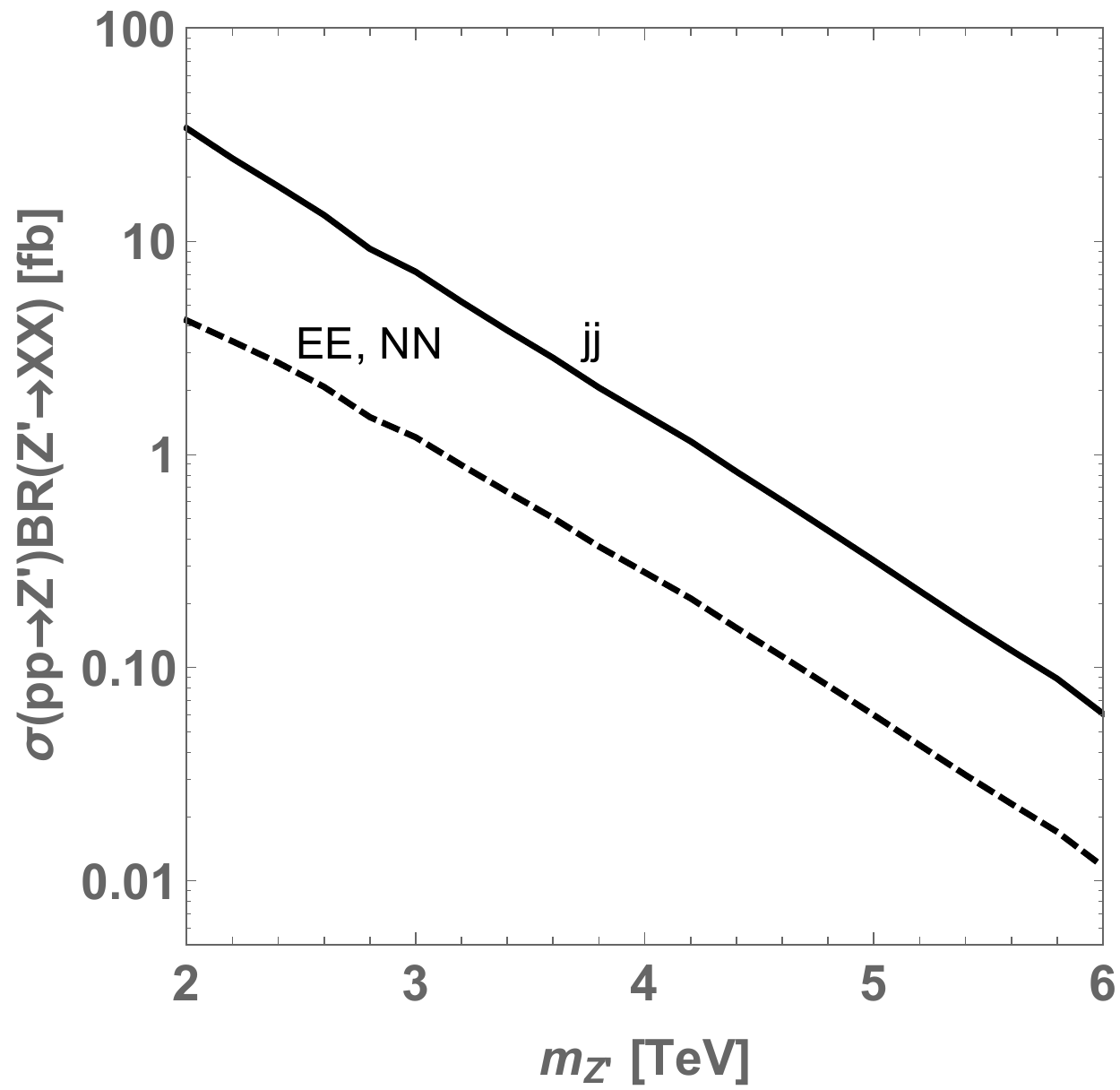} 
\caption{The products of $Z'$ production cross section and the BRs for $jj$ and $E\bar{E}(N \bar{N})$ modes at the LHC 14 TeV where we chose $m_{E(N)} = 700$ GeV for three generations. }
  \label{fig:CX}
\end{center}\end{figure}
 
 \subsection{Signature of $Z'$ production at collider}
 We consider signature of $Z'$ production at hadron collider experiments focusing on 
 parameter space that can explain W boson mass anomaly.
 In Fig.~\ref{fig:gx-mzp}, we show the $1 \sigma$ region to explain W boson mass anomaly on $\{m_{Z'}, g'\}$ plane where the red colored 
 line indicates the value providing central value of $\Delta T$ in Eq.~\eqref{eq:delT}.
For the parameter region we estimate cross section of $pp \to Z'$ process using {\it CalcHEP}~\cite{Belyaev:2012qa} implementing relevant gauge interactions.
Our $Z'$ dominantly decays into $\{q \bar q, N_R \bar N_R, E_R \bar E_R \}$ modes.
The decay width is estimated by
\begin{equation}
\Gamma (Z' \to f \bar f) = \frac{g'^2 N_c}{12 \pi} m_{Z'} \sqrt{1- \frac{4 m_f^2}{m^2_{Z'}}} \left(1- \frac{m_f^2}{m^2_{Z'}} \right)
\end{equation} 
where $N_c$ is color degrees of freedom.
BRs of $Z'$ decay can be estimated by the width.
In Fig.~\ref{fig:CX}, we show products of cross section and the BRs for $jj$ and $E\bar{E}(N \bar{N})$ modes where we chose $m_{E(N)} = 700$ GeV for three generations. 
We thus obtain sizable cross section that can be tested at future LHC experiments.
Extra fermions $E_a$ and $N_a$ dominantly decay into the SM lepton with boson as
\begin{align}
& E_a \to \ell_i h, \\
& N_a \to \nu_i h, \nu_i Z, \ell_i W,
\end{align} 
where lepton flavor dependence is determined by structure of Yukawa couplings.
Therefore signatures of our $Z'$ are dijet, top quark pair and SM leptons with higgs or gauge bosons.
More detailed numerical analysis is beyond the scope of this paper and left as future work.


\section{Summary and Conclusions}
We have proposed a model with leptophobic $U(1)_R$ gauge symmetry to explain CDF W boson mass anomaly.
The shift of W boson mass is realized via neutral gauge boson mass which induces non-zero oblique $\Delta T$ parameter.
Also we have shown that $U(1)_R$ should be leptophobic since parameter region to explain W bosom mass anomaly 
is excluded by the LHC constraints on $W$ and $Y$ oblique parameters if $Z'$ is coupled to the SM charged leptons.
It is then found that $Z'$ mass and new gauge coupling should satisfy 20 TeV $\lesssim m_{Z'}/g' \lesssim$ 31 TeV 
to accommodate CDF W bosom mass.

In our model charged lepton masses are induced via interactions between the SM leptons and extra charged leptons.
These interactions also induce muon $g-2$ and LFVs at loop level.
We have investigated muon $g-2$ and LFVs taking into current experimental constraints.
It is found that some LFV BRs,  $\mu \to e \gamma$ and $\tau \to \mu \gamma$, tends to be close to current upper limits 
when we explain muon $g-2$ and the BRs can be tested in future experiments.
In addition we have discussed realization of active neutrino mass and mixings via inverse seesaw mechanism.
Moreover we have discussed collider physics focusing on $Z'$ production and its decay at the LHC.
We have found that sizable cross section can be expected considering parameter region explaining W boson mass anomaly. 


\section*{Acknowledgments}
KIN was supported by JSPS Grant-in-Aid for Scientific Research (A) 18H03699, (C) 21K03562, (C) 21K03583, Okayama Foundation for Science and Technology, and Wesco Scientific Promotion Foundation.
This research was supported by an appointment to the JRG Program at the APCTP through the Science and Technology Promotion Fund and Lottery Fund of the Korean Government. This was also supported by the Korean Local Governments - Gyeongsangbuk-do Province and Pohang City (H.O.). 
H. O. is sincerely grateful for KIAS and all the members.
The work was also supported by the Fundamental Research Funds for the Central Universities (T.~N.).


\begin{thebibliography}{99}


\bibitem{CDF:2022hxs}
T.~Aaltonen \textit{et al.} [CDF],
Science \textbf{376} (2022) no.6589, 170-176
doi:10.1126/science.abk1781

\bibitem{ParticleDataGroup:2018ovx}
M.~Tanabashi \textit{et al.} [Particle Data Group],
Phys. Rev. D \textbf{98} (2018) no.3, 030001
doi:10.1103/PhysRevD.98.030001

\bibitem{Fan:2022dck}
Y.~Z.~Fan, T.~P.~Tang, Y.~L.~S.~Tsai and L.~Wu,
[arXiv:2204.03693 [hep-ph]].

\bibitem{Lu:2022bgw}
C.~T.~Lu, L.~Wu, Y.~Wu and B.~Zhu,
[arXiv:2204.03796 [hep-ph]].

\bibitem{Athron:2022qpo}
P.~Athron, A.~Fowlie, C.~T.~Lu, L.~Wu, Y.~Wu and B.~Zhu,
[arXiv:2204.03996 [hep-ph]].

\bibitem{Yuan:2022cpw}
G.~W.~Yuan, L.~Zu, L.~Feng and Y.~F.~Cai,
[arXiv:2204.04183 [hep-ph]].

\bibitem{Strumia:2022qkt}
A.~Strumia,
[arXiv:2204.04191 [hep-ph]].

\bibitem{Yang:2022gvz}
J.~M.~Yang and Y.~Zhang,
[arXiv:2204.04202 [hep-ph]].

\bibitem{deBlas:2022hdk}
J.~de Blas, M.~Pierini, L.~Reina and L.~Silvestrini,
[arXiv:2204.04204 [hep-ph]].

\bibitem{diCortona:2022zjy}
G.~G.~di Cortona and E.~Nardi,
[arXiv:2204.04227 [hep-ph]].

\bibitem{Du:2022pbp}
X.~K.~Du, Z.~Li, F.~Wang and Y.~K.~Zhang,
[arXiv:2204.04286 [hep-ph]].

\bibitem{Tang:2022pxh}
T.~P.~Tang, M.~Abdughani, L.~Feng, Y.~L.~S.~Tsai and Y.~Z.~Fan,
[arXiv:2204.04356 [hep-ph]].

\bibitem{Cacciapaglia:2022xih}
G.~Cacciapaglia and F.~Sannino,
[arXiv:2204.04514 [hep-ph]].

\bibitem{Blennow:2022yfm}
M.~Blennow, P.~Coloma, E.~Fern\'andez-Mart\'\i{}nez and M.~Gonz\'alez-L\'opez,
[arXiv:2204.04559 [hep-ph]].

\bibitem{Sakurai:2022hwh}
K.~Sakurai, F.~Takahashi and W.~Yin,
[arXiv:2204.04770 [hep-ph]].

\bibitem{Fan:2022yly}
J.~Fan, L.~Li, T.~Liu and K.~F.~Lyu,
[arXiv:2204.04805 [hep-ph]].

\bibitem{Liu:2022jdq}
X.~Liu, S.~Y.~Guo, B.~Zhu and Y.~Li,
[arXiv:2204.04834 [hep-ph]].

\bibitem{Lee:2022nqz}
H.~M.~Lee and K.~Yamashita,
[arXiv:2204.05024 [hep-ph]].

\bibitem{Cheng:2022jyi}
Y.~Cheng, X.~G.~He, Z.~L.~Huang and M.~W.~Li,
[arXiv:2204.05031 [hep-ph]].

\bibitem{Song:2022xts}
H.~Song, W.~Su and M.~Zhang,
[arXiv:2204.05085 [hep-ph]].

\bibitem{Bagnaschi:2022whn}
E.~Bagnaschi, J.~Ellis, M.~Madigan, K.~Mimasu, V.~Sanz and T.~You,
[arXiv:2204.05260 [hep-ph]].

\bibitem{Paul:2022dds}
A.~Paul and M.~Valli,
[arXiv:2204.05267 [hep-ph]].

\bibitem{Bahl:2022xzi}
H.~Bahl, J.~Braathen and G.~Weiglein,
[arXiv:2204.05269 [hep-ph]].

\bibitem{Asadi:2022xiy}
P.~Asadi, C.~Cesarotti, K.~Fraser, S.~Homiller and A.~Parikh,
[arXiv:2204.05283 [hep-ph]].

\bibitem{DiLuzio:2022xns}
L.~Di Luzio, R.~Gr\"ober and P.~Paradisi,
[arXiv:2204.05284 [hep-ph]].

\bibitem{Athron:2022isz}
P.~Athron, M.~Bach, D.~H.~J.~Jacob, W.~Kotlarski, D.~St\"ockinger and A.~Voigt,
[arXiv:2204.05285 [hep-ph]].

\bibitem{Gu:2022htv}
J.~Gu, Z.~Liu, T.~Ma and J.~Shu,
[arXiv:2204.05296 [hep-ph]].

\bibitem{Heckman:2022the}
J.~J.~Heckman,
[arXiv:2204.05302 [hep-ph]].

\bibitem{Babu:2022pdn}
K.~S.~Babu, S.~Jana and V.~P.~K.,
[arXiv:2204.05303 [hep-ph]].

\bibitem{Heo:2022dey}
Y.~Heo, D.~W.~Jung and J.~S.~Lee,
[arXiv:2204.05728 [hep-ph]].

\bibitem{Du:2022brr}
X.~K.~Du, Z.~Li, F.~Wang and Y.~K.~Zhang,
[arXiv:2204.05760 [hep-ph]].

\bibitem{Cheung:2022zsb}
K.~Cheung, W.~Y.~Keung and P.~Y.~Tseng,
[arXiv:2204.05942 [hep-ph]].

\bibitem{Crivellin:2022fdf}
A.~Crivellin, M.~Kirk, T.~Kitahara and F.~Mescia,
[arXiv:2204.05962 [hep-ph]].

\bibitem{Endo:2022kiw}
M.~Endo and S.~Mishima,
[arXiv:2204.05965 [hep-ph]].

\bibitem{Biekotter:2022abc}
T.~Biek\"otter, S.~Heinemeyer and G.~Weiglein,
[arXiv:2204.05975 [hep-ph]].

\bibitem{Balkin:2022glu}
R.~Balkin, E.~Madge, T.~Menzo, G.~Perez, Y.~Soreq and J.~Zupan,
[arXiv:2204.05992 [hep-ph]].

\bibitem{Krasnikov:2022xsi}
N.~V.~Krasnikov,
[arXiv:2204.06327 [hep-ph]].

\bibitem{Ahn:2022xeq}
Y.~H.~Ahn, S.~K.~Kang and R.~Ramos,
[arXiv:2204.06485 [hep-ph]].

\bibitem{Han:2022juu}
X.~F.~Han, F.~Wang, L.~Wang, J.~M.~Yang and Y.~Zhang,
[arXiv:2204.06505 [hep-ph]].

\bibitem{Zheng:2022irz}
M.~D.~Zheng, F.~Z.~Chen and H.~H.~Zhang,
[arXiv:2204.06541 [hep-ph]].

\bibitem{Kawamura:2022uft}
J.~Kawamura, S.~Okawa and Y.~Omura,
[arXiv:2204.07022 [hep-ph]].

\bibitem{Ghoshal:2022vzo}
A.~Ghoshal, N.~Okada, S.~Okada, D.~Raut, Q.~Shafi and A.~Thapa,
[arXiv:2204.07138 [hep-ph]].

\bibitem{FileviezPerez:2022lxp}
P.~Fileviez Perez, H.~H.~Patel and A.~D.~Plascencia,
[arXiv:2204.07144 [hep-ph]].

\bibitem{Nagao:2022oin}
K.~I.~Nagao, T.~Nomura and H.~Okada,
[arXiv:2204.07411 [hep-ph]].

\bibitem{Kanemura:2022ahw}
S.~Kanemura and K.~Yagyu,
[arXiv:2204.07511 [hep-ph]].

\bibitem{Mondal:2022xdy}
P.~Mondal,
[arXiv:2204.07844 [hep-ph]].

\bibitem{Zhang:2022nnh}
K.~Y.~Zhang and W.~Z.~Feng,
[arXiv:2204.08067 [hep-ph]].

\bibitem{Borah:2022obi}
D.~Borah, S.~Mahapatra, D.~Nanda and N.~Sahu,
[arXiv:2204.08266 [hep-ph]].

\bibitem{Chowdhury:2022moc}
T.~A.~Chowdhury, J.~Heeck, S.~Saad and A.~Thapa,
[arXiv:2204.08390 [hep-ph]].

\bibitem{Arcadi:2022dmt}
G.~Arcadi and A.~Djouadi,
[arXiv:2204.08406 [hep-ph]].

\bibitem{Cirigliano:2022qdm}
V.~Cirigliano, W.~Dekens, J.~de Vries, E.~Mereghetti and T.~Tong,
[arXiv:2204.08440 [hep-ph]].

\bibitem{Carpenter:2022oyg}
L.~M.~Carpenter, T.~Murphy and M.~J.~Smylie,
[arXiv:2204.08546 [hep-ph]].

\bibitem{Popov:2022ldh}
O.~Popov and R.~Srivastava,
[arXiv:2204.08568 [hep-ph]].

\bibitem{Ghorbani:2022vtv}
K.~Ghorbani and P.~Ghorbani,
[arXiv:2204.09001 [hep-ph]].

\bibitem{Du:2022fqv}
M.~Du, Z.~Liu and P.~Nath,
[arXiv:2204.09024 [hep-ph]].

\bibitem{Bhaskar:2022vgk}
A.~Bhaskar, A.~A.~Madathil, T.~Mandal and S.~Mitra,
[arXiv:2204.09031 [hep-ph]].

\bibitem{Batra:2022org}
A.~Batra, S.~K.A., S.~Mandal and R.~Srivastava,
[arXiv:2204.09376 [hep-ph]].

\bibitem{Cao:2022mif}
J.~Cao, L.~Meng, L.~Shang, S.~Wang and B.~Yang,
[arXiv:2204.09477 [hep-ph]].


\bibitem{Zeng:2022lkk}
Y.~P.~Zeng, C.~Cai, Y.~H.~Su and H.~H.~Zhang,
[arXiv:2204.09487 [hep-ph]].

\bibitem{Baek:2022agi}
S.~Baek,
[arXiv:2204.09585 [hep-ph]].

\bibitem{Borah:2022zim}
D.~Borah, S.~Mahapatra and N.~Sahu,
[arXiv:2204.09671 [hep-ph]].

\bibitem{Almeida:2022lcs}
E.~d.~Almeida, A.~Alves, O.~J.~P.~Eboli and M.~C.~Gonzalez-Garcia,
[arXiv:2204.10130 [hep-ph]].

\bibitem{Cheng:2022aau}
Y.~Cheng, X.~G.~He, F.~Huang, J.~Sun and Z.~P.~Xing,
[arXiv:2204.10156 [hep-ph]].

\bibitem{Heeck:2022fvl}
J.~Heeck,
[arXiv:2204.10274 [hep-ph]].

\bibitem{Addazi:2022fbj}
A.~Addazi, A.~Marciano, A.~P.~Morais, R.~Pasechnik and H.~Yang,
[arXiv:2204.10315 [hep-ph]].

\bibitem{Lee:2022gyf}
S.~Lee, K.~Cheung, J.~Kim, C.~T.~Lu and J.~Song,
[arXiv:2204.10338 [hep-ph]].

\bibitem{Cai:2022cti}
C.~Cai, D.~Qiu, Y.~L.~Tang, Z.~H.~Yu and H.~H.~Zhang,
[arXiv:2204.11570 [hep-ph]].

\bibitem{Benbrik:2022dja}
R.~Benbrik, M.~Boukidi and B.~Manaut,
[arXiv:2204.11755 [hep-ph]].

\bibitem{Yang:2022qgs}
T.~Yang, S.~Qian, S.~Deng, J.~Xiao, L.~Gao, A.~M.~Levin, Q.~Li, M.~Lu and Z.~You,
[arXiv:2204.11871 [hep-ph]].

\bibitem{Batra:2022pej}
A.~Batra, S.~K.~A, S.~Mandal, H.~Prajapati and R.~Srivastava,
[arXiv:2204.11945 [hep-ph]].

\bibitem{Tan:2022bip}
H.~B.~T.~Tan and A.~Derevianko,
[arXiv:2204.11991 [hep-ph]].

\bibitem{Abouabid:2022lpg}
H.~Abouabid, A.~Arhrib, R.~Benbrik, M.~Krab and M.~Ouchemhou,
[arXiv:2204.12018 [hep-ph]].

\bibitem{Chen:2022ocr}
T.~K.~Chen, C.~W.~Chiang and K.~Yagyu,
[arXiv:2204.12898 [hep-ph]].

\bibitem{Zhou:2022cql}
Q.~Zhou and X.~F.~Han,
[arXiv:2204.13027 [hep-ph]].

\bibitem{Gupta:2022lrt}
R.~S.~Gupta,
[arXiv:2204.13690 [hep-ph]].

\bibitem{Basiouris:2022wei}
V.~Basiouris and G.~K.~Leontaris,
[arXiv:2205.00758 [hep-ph]].

\bibitem{Wang:2022dte}
J.~W.~Wang, X.~J.~Bi, P.~F.~Yin and Z.~H.~Yu,
[arXiv:2205.00783 [hep-ph]].

\bibitem{Botella:2022rte}
F.~J.~Botella, F.~Cornet-Gomez, C.~Mir\'o and M.~Nebot,
[arXiv:2205.01115 [hep-ph]].

\bibitem{Kim:2022xuo}
J.~Kim,
Phys. Lett. B \textbf{832} (2022), 137220
[arXiv:2205.01437 [hep-ph]].
Copy to ClipboardDownload


\bibitem{Barman:2022qix}
B.~Barman, A.~Das and S.~Sengupta,
[arXiv:2205.01699 [hep-ph]].

\bibitem{Kim:2022hvh}
J.~Kim, S.~Lee, P.~Sanyal and J.~Song,
[arXiv:2205.01701 [hep-ph]].

\bibitem{Li:2022gwc}
X.~Q.~Li, Z.~J.~Xie, Y.~D.~Yang and X.~B.~Yuan,
[arXiv:2205.02205 [hep-ph]].

\bibitem{Isaacson:2022rts}
J.~Isaacson, Y.~Fu and C.~P.~Yuan,
[arXiv:2205.02788 [hep-ph]].

\bibitem{Evans:2022dgq}
J.~L.~Evans, T.~T.~Yanagida and N.~Yokozaki,
[arXiv:2205.03877 [hep-ph]].

\bibitem{Chowdhury:2022dps}
T.~A.~Chowdhury and S.~Saad,
[arXiv:2205.03917 [hep-ph]].

\bibitem{Kim:2022zhj}
S.~S.~Kim, H.~M.~Lee, A.~G.~Menkara and K.~Yamashita,
[arXiv:2205.04016 [hep-ph]].

\bibitem{Lazarides:2022spe}
G.~Lazarides, R.~Maji, R.~Roshan and Q.~Shafi,
[arXiv:2205.04824 [hep-ph]].

\bibitem{Senjanovic:2022zwy}
G.~Senjanovi\'c and M.~Zantedeschi,
[arXiv:2205.05022 [hep-ph]].

\bibitem{Ghosh:2022zqs}
R.~Ghosh, B.~Mukhopadhyaya and U.~Sarkar,
[arXiv:2205.05041 [hep-ph]].

\bibitem{Li:2022eby}
T.~Li, J.~Pei, X.~Yin and B.~Zhu,
[arXiv:2205.08215 [hep-ph]].

\bibitem{Rodriguez:2022wix}
M.~C.~Rodriguez,
[arXiv:2205.09109 [hep-ph]].

\bibitem{Ramirez:2022zpk}
E.~Ramirez and P.~Roig,
[arXiv:2205.10420 [hep-ph]].

\bibitem{Kawamura:2022fhm}
J.~Kawamura and S.~Raby,
[arXiv:2205.10480 [hep-ph]].

\bibitem{Afonin:2022cbi}
S.~S.~Afonin,
[arXiv:2205.12237 [hep-ph]].

\bibitem{Allanach:2022bik}
B.~C.~Allanach and J.~Davighi,
[arXiv:2205.12252 [hep-ph]].



\bibitem{Xue:2022mde}
S.~S.~Xue,
[arXiv:2205.14957 [hep-ph]].

\bibitem{Rizzo:2022jti}
T.~G.~Rizzo,
[arXiv:2206.09814 [hep-ph]].

\bibitem{VanLoi:2022eir}
D.~Van Loi and P.~Van Dong,
[arXiv:2206.10100 [hep-ph]].

\bibitem{YaserAyazi:2022tbn}
S.~Yaser Ayazi and M.~Hosseini,
[arXiv:2206.11041 [hep-ph]].

\bibitem{Chakrabarty:2022voz}
N.~Chakrabarty,
[arXiv:2206.11771 [hep-ph]].

\bibitem{CentellesChulia:2022vpz}
S.~Centelles Chuli\'a, R.~Srivastava and S.~Yadav,
[arXiv:2206.11903 [hep-ph]].


\bibitem{Peskin:1991sw}
M.~E.~Peskin and T.~Takeuchi,
Phys. Rev. D \textbf{46} (1992), 381-409
doi:10.1103/PhysRevD.46.381

\bibitem{Peskin:1990zt}
M.~E.~Peskin and T.~Takeuchi,
Phys. Rev. Lett. \textbf{65} (1990), 964-967
doi:10.1103/PhysRevLett.65.964



\bibitem{Jung:2009jz} 
  S.~Jung, H.~Murayama, A.~Pierce and J.~D.~Wells,
  Phys.\ Rev.\ D {\bf 81}, 015004 (2010)
  [arXiv:0907.4112 [hep-ph]].



\bibitem{Nomura:2017tih}
T.~Nomura and H.~Okada,
Phys. Rev. D \textbf{97} (2018) no.1, 015015
doi:10.1103/PhysRevD.97.015015
[arXiv:1707.00929 [hep-ph]].


  
\bibitem{Nomura:2017abh} 
  T.~Nomura, H.~Okada and H.~Yokoya,
  arXiv:1702.03396 [hep-ph].

\bibitem{Nomura:2018mwr}
T.~Nomura and H.~Okada,
LHEP \textbf{1} (2018) no.2, 10-13
doi:10.31526/LHEP.2.2018.01
[arXiv:1806.01714 [hep-ph]].



\bibitem{Nomura:2017ezy} 
  T.~Nomura and H.~Okada,
  arXiv:1704.03382 [hep-ph].
  
    
\bibitem{Chao:2017rwv} 
  W.~Chao,
  arXiv:1707.07858 [hep-ph].
  
\bibitem{Seto:2020jal}
O.~Seto and T.~Shimomura,
JHEP \textbf{04} (2021), 025
[arXiv:2006.05497 [hep-ph]].

  
\bibitem{Lee:2021gnw}
H.~M.~Lee, J.~Song and K.~Yamashita,
J. Korean Phys. Soc. \textbf{79} (2021) no.12, 1121-1134
[arXiv:2110.09942 [hep-ph]].


\bibitem{Guedes:2022cfy}
G.~Guedes and P.~Olgoso,
[arXiv:2205.04480 [hep-ph]].

  
  
\bibitem{Dey:2019cts}
U.~K.~Dey, T.~Nomura and H.~Okada,
Phys. Rev. D \textbf{100} (2019) no.7, 075013
doi:10.1103/PhysRevD.100.075013
[arXiv:1902.06205 [hep-ph]].

  

\bibitem{Esteban:2018azc}
I.~Esteban, M.~C.~Gonzalez-Garcia, A.~Hernandez-Cabezudo, M.~Maltoni and T.~Schwetz,
JHEP \textbf{01} (2019), 106
[arXiv:1811.05487 [hep-ph]].

\bibitem{Fernandez-Martinez:2016lgt}
E.~Fernandez-Martinez, J.~Hernandez-Garcia and J.~Lopez-Pavon,
JHEP \textbf{08} (2016), 033
doi:10.1007/JHEP08(2016)033
[arXiv:1605.08774 [hep-ph]].

\bibitem{Agostinho:2017wfs}
N.~R.~Agostinho, G.~C.~Branco, P.~M.~F.~Pereira, M.~N.~Rebelo and J.~I.~Silva-Marcos,
Eur. Phys. J. C \textbf{78} (2018) no.11, 895
doi:10.1140/epjc/s10052-018-6347-2
[arXiv:1711.06229 [hep-ph]].

  
  
  \bibitem{Belyaev:2012qa} 
  A.~Belyaev, N.~D.~Christensen and A.~Pukhov,
  Comput.\ Phys.\ Commun.\  {\bf 184}, 1729 (2013)
  [arXiv:1207.6082 [hep-ph]].
  
\bibitem{Nadolsky:2008zw} 
  P.~M.~Nadolsky, H.~L.~Lai, Q.~H.~Cao, J.~Huston, J.~Pumplin, D.~Stump, W.~K.~Tung and C.-P.~Yuan,
  Phys.\ Rev.\ D {\bf 78}, 013004 (2008)
  [arXiv:0802.0007 [hep-ph]].
  


  
\bibitem{Cerrito:2016qig} 
  L.~Cerrito, D.~Millar, S.~Moretti and F.~Spano,
  arXiv:1609.05540 [hep-ph].
  
  
    
  
   \bibitem{pdg} P.~A.~Zyla {\it et al.} (Particle Data Group), Prog. Theor. Exp. Phys. {\bf 2020}, 083C01 (2020).
   
   
   
\bibitem{ALEPH:2013dgf}
S.~Schael \textit{et al.} [ALEPH, DELPHI, L3, OPAL and LEP Electroweak],
Phys. Rept. \textbf{532} (2013), 119-244
[arXiv:1302.3415 [hep-ex]].
   
   
\bibitem{ATLAS:2019fgd}
G.~Aad \textit{et al.} [ATLAS],
JHEP \textbf{03} (2020), 145
[arXiv:1910.08447 [hep-ex]].

   
\bibitem{CMS:2019gwf}
A.~M.~Sirunyan \textit{et al.} [CMS],
JHEP \textbf{05} (2020), 033
[arXiv:1911.03947 [hep-ex]].



\bibitem{Dobrescu:2021vak}
B.~A.~Dobrescu and F.~Yu,
[arXiv:2112.05392 [hep-ph]].

\bibitem{Cacciapaglia:2006pk}
G.~Cacciapaglia, C.~Csaki, G.~Marandella and A.~Strumia,
Phys. Rev. D \textbf{74} (2006), 033011
doi:10.1103/PhysRevD.74.033011
[arXiv:hep-ph/0604111 [hep-ph]].

\bibitem{CMS:2022yjm}
A.~Tumasyan \textit{et al.} [CMS],
JHEP \textbf{07} (2022), 067
doi:10.1007/JHEP07(2022)067
[arXiv:2202.06075 [hep-ex]].

\bibitem{ATLAS:2017fih}
M.~Aaboud \textit{et al.} [ATLAS],
JHEP \textbf{10} (2017), 182
doi:10.1007/JHEP10(2017)182
[arXiv:1707.02424 [hep-ex]].
   
   
   

   
   
   
  
\bibitem{MEG:2016leq}
A.~M.~Baldini \textit{et al.} [MEG],
Eur. Phys. J. C \textbf{76} (2016) no.8, 434
doi:10.1140/epjc/s10052-016-4271-x
[arXiv:1605.05081 [hep-ex]].

   
\bibitem{MEG:2013oxv}
J.~Adam \textit{et al.} [MEG],
Phys. Rev. Lett. \textbf{110} (2013), 201801
doi:10.1103/PhysRevLett.110.201801
[arXiv:1303.0754 [hep-ex]].


  

\bibitem{Muong-2:2021ojo}
B.~Abi \textit{et al.} [Muon g-2],
Phys. Rev. Lett. \textbf{126} (2021) no.14, 141801
doi:10.1103/PhysRevLett.126.141801
[arXiv:2104.03281 [hep-ex]].


\bibitem{Hagiwara:2011af} 
  K.~Hagiwara, R.~Liao, A.~D.~Martin, D.~Nomura and T.~Teubner,
  J.\ Phys.\ G {\bf 38}, 085003 (2011)
  [arXiv:1105.3149 [hep-ph]].
  

\bibitem{Aoyama:2012wk}
T.~Aoyama, M.~Hayakawa, T.~Kinoshita and M.~Nio,
Phys. Rev. Lett. \textbf{109}, 111808 (2012)
doi:10.1103/PhysRevLett.109.111808
[arXiv:1205.5370 [hep-ph]].

\bibitem{Aoyama:2019ryr}
T.~Aoyama, T.~Kinoshita and M.~Nio,
Atoms \textbf{7}, no.1, 28 (2019)
doi:10.3390/atoms7010028

\bibitem{Czarnecki:2002nt}
A.~Czarnecki, W.~J.~Marciano and A.~Vainshtein,
Phys. Rev. D \textbf{67}, 073006 (2003)
[erratum: Phys. Rev. D \textbf{73}, 119901 (2006)]
doi:10.1103/PhysRevD.67.073006
[arXiv:hep-ph/0212229 [hep-ph]].

\bibitem{Gnendiger:2013pva}
C.~Gnendiger, D.~St\"ockinger and H.~St\"ockinger-Kim,
Phys. Rev. D \textbf{88}, 053005 (2013)
doi:10.1103/PhysRevD.88.053005
[arXiv:1306.5546 [hep-ph]].

\bibitem{Keshavarzi:2018mgv}
A.~Keshavarzi, D.~Nomura and T.~Teubner,
Phys. Rev. D \textbf{97}, no.11, 114025 (2018)
doi:10.1103/PhysRevD.97.114025
[arXiv:1802.02995 [hep-ph]].

\bibitem{Colangelo:2018mtw}
G.~Colangelo, M.~Hoferichter and P.~Stoffer,
JHEP \textbf{02}, 006 (2019)
doi:10.1007/JHEP02(2019)006
[arXiv:1810.00007 [hep-ph]].

\bibitem{Hoferichter:2019mqg}
M.~Hoferichter, B.~L.~Hoid and B.~Kubis,
JHEP \textbf{08}, 137 (2019)
doi:10.1007/JHEP08(2019)137
[arXiv:1907.01556 [hep-ph]].

\bibitem{Keshavarzi:2019abf}
A.~Keshavarzi, D.~Nomura and T.~Teubner,
Phys. Rev. D \textbf{101}, no.1, 014029 (2020)
doi:10.1103/PhysRevD.101.014029
[arXiv:1911.00367 [hep-ph]].

\bibitem{Kurz:2014wya}
A.~Kurz, T.~Liu, P.~Marquard and M.~Steinhauser,
Phys. Lett. B \textbf{734}, 144-147 (2014)
doi:10.1016/j.physletb.2014.05.043
[arXiv:1403.6400 [hep-ph]].

\bibitem{Melnikov:2003xd}
K.~Melnikov and A.~Vainshtein,
Phys. Rev. D \textbf{70}, 113006 (2004)
doi:10.1103/PhysRevD.70.113006
[arXiv:hep-ph/0312226 [hep-ph]].

\bibitem{Masjuan:2017tvw}
P.~Masjuan and P.~Sanchez-Puertas,
Phys. Rev. D \textbf{95}, no.5, 054026 (2017)
doi:10.1103/PhysRevD.95.054026
[arXiv:1701.05829 [hep-ph]].

\bibitem{Colangelo:2017fiz}
G.~Colangelo, M.~Hoferichter, M.~Procura and P.~Stoffer,
JHEP \textbf{04}, 161 (2017)
doi:10.1007/JHEP04(2017)161
[arXiv:1702.07347 [hep-ph]].

\bibitem{Hoferichter:2018kwz}
M.~Hoferichter, B.~L.~Hoid, B.~Kubis, S.~Leupold and S.~P.~Schneider,
JHEP \textbf{10}, 141 (2018)
doi:10.1007/JHEP10(2018)141
[arXiv:1808.04823 [hep-ph]].

\bibitem{Gerardin:2019vio}
A.~G\'erardin, H.~B.~Meyer and A.~Nyffeler,
Phys. Rev. D \textbf{100}, no.3, 034520 (2019)
doi:10.1103/PhysRevD.100.034520
[arXiv:1903.09471 [hep-lat]].

\bibitem{Bijnens:2019ghy}
J.~Bijnens, N.~Hermansson-Truedsson and A.~Rodr\'\i{}guez-S\'anchez,
Phys. Lett. B \textbf{798}, 134994 (2019)
doi:10.1016/j.physletb.2019.134994
[arXiv:1908.03331 [hep-ph]].

\bibitem{Colangelo:2019uex}
G.~Colangelo, F.~Hagelstein, M.~Hoferichter, L.~Laub and P.~Stoffer,
JHEP \textbf{03}, 101 (2020)
doi:10.1007/JHEP03(2020)101
[arXiv:1910.13432 [hep-ph]].

\bibitem{Blum:2019ugy}
T.~Blum, N.~Christ, M.~Hayakawa, T.~Izubuchi, L.~Jin, C.~Jung and C.~Lehner,
Phys. Rev. Lett. \textbf{124}, no.13, 132002 (2020)
doi:10.1103/PhysRevLett.124.132002
[arXiv:1911.08123 [hep-lat]].

\bibitem{Colangelo:2014qya}
G.~Colangelo, M.~Hoferichter, A.~Nyffeler, M.~Passera and P.~Stoffer,
Phys. Lett. B \textbf{735}, 90-91 (2014)
doi:10.1016/j.physletb.2014.06.012
[arXiv:1403.7512 [hep-ph]].


\bibitem{Davier:2017zfy}
M.~Davier, A.~Hoecker, B.~Malaescu and Z.~Zhang,
Eur. Phys. J. C \textbf{77}, no.12, 827 (2017)
doi:10.1140/epjc/s10052-017-5161-6
[arXiv:1706.09436 [hep-ph]].

\bibitem{Davier:2019can}
M.~Davier, A.~Hoecker, B.~Malaescu and Z.~Zhang,
Eur. Phys. J. C \textbf{80}, no.3, 241 (2020)
[erratum: Eur. Phys. J. C \textbf{80}, no.5, 410 (2020)]
doi:10.1140/epjc/s10052-020-7792-2
[arXiv:1908.00921 [hep-ph]].

   
   

    
\end{thebibliography}
\end{document}